%% file: main.tex
\documentclass[onecolumn,12pt,tightenlines,superscriptaddress,notitlepage]{revtex4-1}

\usepackage{graphicx}
\usepackage{amsmath}
\usepackage{amssymb}
\usepackage[titletoc]{appendix}
 
\usepackage{bm}
\usepackage{amsfonts}
\usepackage{epstopdf}
\usepackage{xcolor}
\usepackage{color}
\usepackage{wasysym}
\usepackage{hyperref}
\usepackage{subcaption}

\numberwithin{equation}{section}

\hypersetup{
    colorlinks,    citecolor=blue,    filecolor=blue,    linkcolor=blue,    urlcolor=blue
}

\usepackage{verbatim}

\date{\today}

\begin{document}

\begin{abstract}
While plasma often behaves diamagnetically, we demonstrate that the laser irradiation of a thin opaque target with an embedded target-transverse seed magnetic field $B_\mathrm{seed}$ can trigger the generation of an order-of-magnitude stronger magnetic field with opposite sign at the target surface. 
Strong surface field generation occurs when the laser pulse is relativistically intense and results from the currents associated with the cyclotron rotation of laser-heated electrons transiting through the target and the compensating current of cold electrons.
We derive a predictive scaling for this surface field generation, $B_\mathrm{gen} \sim - 2 \pi B_\mathrm{seed} \Delta x/\lambda_0$, where $\Delta x$ is the target thickness and $\lambda_0$ is the laser wavelength, and conduct 1D and 2D particle-in-cell simulations to confirm its applicability over a wide range of conditions.
We additionally demonstrate that both the seed and surface-generated magnetic fields can have a strong impact on application-relevant plasma dynamics, for example substantially altering the overall expansion and ion acceleration from a $\mu$m-thick laser-irradiated target with a kilotesla-level seed magnetic field.

\end{abstract}

\title{Strong surface magnetic field generation in relativistic short pulse laser-plasma interaction with an applied seed magnetic field}

\author{K. Weichman}
\affiliation{Department of Mechanical and Aerospace Engineering, University of California at San Diego, La Jolla, CA 92093, USA}
\author{A.P.L. Robinson}
\affiliation{Central Laser Facility, STFC Rutherford-Appleton Laboratory, Didcot, OX11 0QX, UK}
\author{M. Murakami}
\affiliation{Institute of Laser Engineering, Osaka University, Suita, Osaka 565-0871, Japan}
\author{A.V. Arefiev}
\affiliation{Department of Mechanical and Aerospace Engineering, University of California at San Diego, La Jolla, CA 92093, USA}
\affiliation{Center for Energy Research, University of California at San Diego, La Jolla, CA 92037, USA}

\vskip -2.0cm
\maketitle
\vskip -3.0cm

\section{Introduction}

Relativisitic laser-plasma interaction with applied magnetic fields presents an opportunity to study the effects of magnetic fields in the high energy density regime. 
Both applied and self-generated magnetic fields can strongly influence plasma behavior, and make laser-plasma a convenient platform both for investigating the fundamental physics of magnetized plasmas, for example laboratory astrophysics~\cite{huntington2015weibel,fiskel2014reconnection,bulanov2015labastro}, and for exploring potential improvements to laser-plasma applications, such as inertial fusion energy~\cite{strozzi2012fast_ignition,fujioka2016fast_ignition,sakata2018isochoric}.

Plasma has a reputation for being diamagnetic and often acts to exclude magnetic fields. However, in the laser-plasma context, there is growing interest in scenarios where laser-plasma interactions have the potential to self-generate strong magnetic fields or to amplify weak applied magnetic fields~\cite{sheng1996inversefaraday,borghesi1998azimuthal,robinson2014transport,huang2019bgen_solid,gotchev2009fluxcompression,meinecke2014turbulent}.
Such an objective is desirable to augment experimentally available magnetic fields from laser-driven coil~\cite{fujioka2013coil,santos2015coil,gao2016coil,goyon2017coil}
or pulsed power sources~\cite{portugall1997field_gen,portugall1999field_gen,ivanov2018zebra} and push the study of magnetized high energy density physics into new regimes.
Most of the previous work has relied on instability-seeded growth~\cite{meinecke2014turbulent,huntington2015weibel}, flux compression~\cite{gotchev2009fluxcompression}, or circularly polarized or Laguerre-Gaussian~\cite{sheng1996inversefaraday,ali2010inversefaraday} laser pulses, which limits these laser-driven magnetic field generation techniques to specific experimental facilities. However, it has recently been shown that the more ubiquitous Gaussian linearly polarized laser pulses also have the potential to amplify a target-normal seed magnetic field in a thin overdense (i.e. opaque) target~\cite{shi2020amp}. 

In this work, we demonstrate that an embedded target-transverse magnetic field can also trigger the generation of a strong surface magnetic field. 
We find that the generation of a non-azimuthal large-amplitude magnitude field at the rear target surface results from the localized production of electrons at the laser-irradiated surface and requires relativistic laser intensity.
We additionally construct a predictive scaling based on the physical processes driving the magnetic field generation.
This scaling is robust over a wide range of laser and target conditions. 

We further demonstrate the validity of our predictive scaling and the importance of surface magnetic field generation in applications. 
As an example, we consider the effect of the seed and surface-generated magnetic fields on the dynamics of target expansion and ion acceleration from a laser-irradiated target.
Both the applied and plasma-generated fields can become sufficiently large to modify ion acceleration from the target surfaces. 
As we will demonstrate, the surface-generated magnetic field can become sufficiently strong to restrict the expansion of the rear target surface. Meanwhile, the seed field can facilitate ion acceleration from the laser-irradiated surface, in some cases even causing the front-surface acceleration to outperform the rear-surface acceleration.

The outline of this paper is as follows. In Section~\ref{sec:1Ddemo}, we conduct 1D simulations and demonstrate that strong surface field generation is tied to cyclotron rotation of the hot and cold electron return currents generated by laser-plasma interaction in an embedded magnetic field. In Section~\ref{sec:1Destimates}, we estimate the magnitude of the rear surface field and show that the surface field generation is robust over a wide parameter range.
In Section~\ref{sec:trapping}, we demonstrate with 1D and 2D simulations that the generation of strong surface fields can initiate electron confinement near the target surfaces and that this confinement can strongly impact the expansion and acceleration of ions from a laser-irradiated target.
In Section~\ref{sec:summary}, we summarize and discuss possible extensions of this work.


\section{Surface magnetic field generation} \label{sec:1Ddemo}

In this Section, we will discuss how laser-irradiation of an opaque target with an embedded target-transverse magnetic field is able to produce strong surface magnetic fields. 
We will initially demonstrate this using 1D particle-in-cell simulations.

\begin{figure*}
    \centering
    \includegraphics[width=0.95\linewidth]{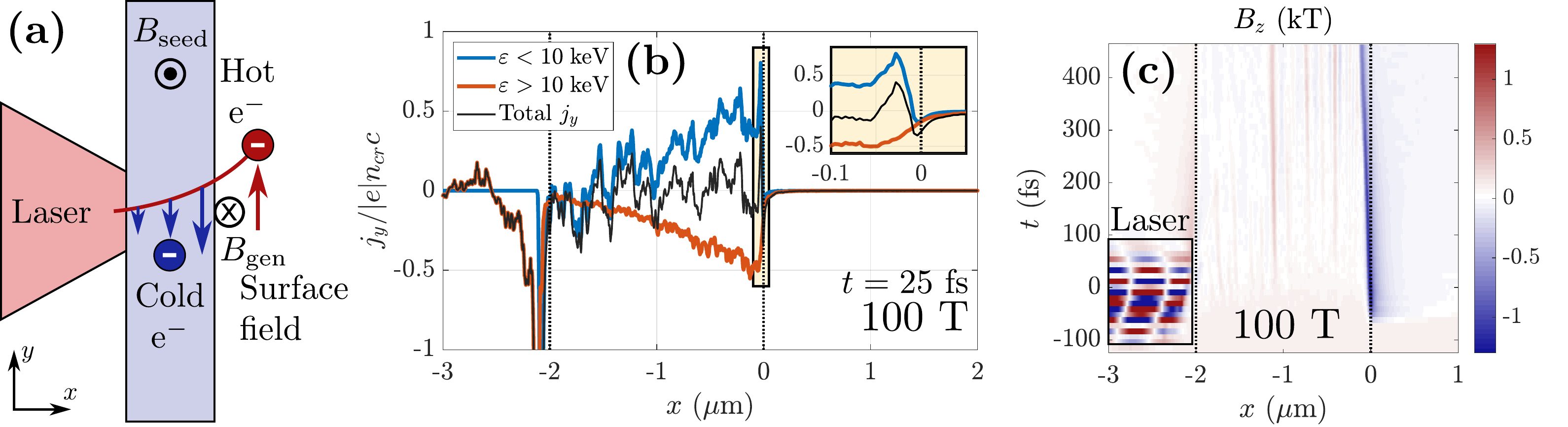}
    \caption{Surface magnetic field generation in laser-irradiated targets. (a) Hot electrons streaming through the target gain transverse momentum in the presence of $B_{\mathrm{seed}}$, inducing a counter-streaming current in the cold target population and generating a surface magnetic field. 
    (b) Current density $j_y$ generated by electrons in 1D PIC simulation, $B_{\mathrm{seed}}=100$~T. Inset: current near the target rear surface.
    The spike near the front surface is associated with electron motion in the laser and does not contribute significantly to the time-averaged magnetic field generation.
    (c) Surface-tangent magnetic field $B_z$ generated in 1D PIC simulation with $B_{\mathrm{seed}}=100$~T. The pattern associated with the laser at the front surface (black box) is an artifact of the time averaging (see Table~\ref{table:1DPIC}) and the data output frequency. Dotted lines in (b) and (c) indicate the initial target surfaces.
    }
    \label{fig:1D_jdiv}
\end{figure*}

We simulate a relativisitcally intense laser pulse interacting with a plastic (CH) target with an embedded target-transverse magnetic field. We conduct collisionless simulations using the open-source particle-in-cell code EPOCH~\cite{arber2015epoch}. The target is represented by a quasineutral CH plasma with a short scale length preplasma and peak density $n_e = 50\;n_{cr}$, where $n_{c} = 1.75\times 10^{21}$~cm$^{-3}$ is the critical density corresponding to the laser wavelength.
The simulation parameters for our nominal case are given in Table \ref{table:1DPIC}. The simulation setup is shown schematically in Fig.~\ref{fig:1D_jdiv}.

\begin{table}
\centering
\begin{tabular}{ |l|l| }
  \hline
  \multicolumn{2}{|l|}{\textbf{Laser parameters} }\\
  \hline
  Wavelength & $\lambda_0=0.8$ $\mu$m \\
  Peak intensity & $1 \times 10^{19}$ W/cm$^2$ \\
  Duration (Gaussian, electric field FWHM) & $100$ fs\\
  Laser propagation direction & $+x$ \\
  Laser polarization & $y$ \\
  \hline \hline
  \multicolumn{2}{|l|}{\textbf{Other parameters} }\\
  \hline
  Seed magnetic field ($\mathbf{B}=B_{\mathrm{seed}} \mathbf{\hat{z}}$) & $B_{\mathrm{seed}}=100$ T \\  
  Target thickness & $\Delta x = 2$ $\mu$m\\
  Preplasma scale length (1/e dropoff) & 0.1 $\mu$m \\
  Peak electron density & $n_e = 50\;n_{cr}$ \\  
  Preplasma density cutoff (minimum) & $0.05\;n_{cr}$ \\  
  Spatial resolution & 200 cells/$\lambda_0$ \\
  Macroparticles per cell, electron &  400 \\
  Macroparticles per cell, ion &  200 \\  
  Time interval for averaging $B_z$ in figures & 10 fs \\
  \hline \hline
  \multicolumn{2}{|l|}{\textbf{Position and time reference}} \\
  \hline
  Location of the front of the foil & $x = 0$ \\
  Time when peak of laser would reach $x=0$ & $t = 0$ \\
  \hline
   \end{tabular}
  \caption{Nominal 1D PIC simulation parameters. The initial plasma temperature is set as zero. The simulation setup is shown schematically in Fig.~\ref{fig:1D_jdiv}a.}
  \label{table:1DPIC}
\end{table}

We observe the generation of surface magnetic fields with 10-15 times the magnitude of the original seed (for example, Fig.~\ref{fig:1D_jdiv}c).
In 1D simulations, the strongest field is generated at the rear target surface ($x=0$) and a weaker field of opposite sign is generated near the laser-irradiated surface ($x=-2$~$\mu$m). These fields rise quickly (on the order of the 100~fs pulse duration) and persist for hundreds of femtoseconds after the laser pulse has been fully reflected by the target. In the rest of this work, we will focus primarily on the rear-surface field. Unlike the front surface field, the rear surface field can have high amplitude and is also present in higher dimensional simulations (e.g. Sec.~\ref{sec:trapping}).

The rear-surface field is produced as a consequence of the cyclotron rotation of the laser-heated and cold (return current) electrons propagating through the target. 
As shown both schematically and quantitatively in Fig.~\ref{fig:1D_jdiv}, the current which creates this field can be separated into contributions from the hot and cold electron populations. 

We choose the division between hot and cold in Fig.~\ref{fig:1D_jdiv} to be 10~keV to fully capture the contribution of each population. However, in spite of this seemingly low energy, we do not expect collisions (which are not included in our simulations) to disrupt the surface magnetic field generation. The small angle collisional scattering time ($1/\nu_{ei}$) is approximately
\begin{equation}
\begin{split}
    1/\nu_{ei} & \approx \dfrac{m_e^2 v_0^3}{8 \pi Z^2 e^4 n_i} \left(\ln\dfrac{m_e v_0^2}{2 Z e^2 n_i^{1/3}}\right)^{-1} \\
    & = 1.3\, \mathrm{ns} \cdot \dfrac{(v_0/c)^3}{Z^2} \left(11+\ln\dfrac{(v_0/c)^2}{Z}\right)^{-1},
\end{split}    
\end{equation}
where $v_0$ is the velocity of hot electrons, $Z$ is the charge state of the ions, in the Coulomb logarithm ($\ln r_\mathrm{max}/r_\mathrm{min}$) we have approximated $r_\mathrm{max}$ by the ion spacing $n_i^{-1/3}$, and in the second expression we have used $n_i = n_e/7 = 1.25\times 10^{22}$~cm$^{-3}$.
The majority of the hot electron current is carried by electrons with energy above 25~keV ($v_0/c \sim 0.3$), which corresponds to a small angle collisional scattering time for carbon ions ($Z=6$) of approximately 150~fs. This is much longer than both the transit time of the electron through the target and the rise time of the surface magnetic field (both $\sim 30$~fs).

The transverse currents responsible for the surface magnetic field generation are driven by the travel of hot electrons through the magnetized target.
Hot electrons are generated at the front target surface by the interaction of the laser with the preplasma. 
These electrons then stream through the target with a net $+x$-directed velocity, during which time the embedded $+z$-directed seed magnetic field $B_{\mathrm{seed}}$ rotates their momentum such that they exit the rear target surface carrying a net transverse current $j_y < 0$ (red line in Fig.~\ref{fig:1D_jdiv}b). 
In response, the cold electrons in the target obtain a compensating $j_y > 0$ which prevents the embedded magnetic field from decreasing (blue line in Fig.~\ref{fig:1D_jdiv}b). However, only hot electrons are able to enter the rear target sheath, resulting in an uncompensated current in the sheath (hot electrons) and in response at the rear surface (cold electrons), as shown in the inset in Fig.~\ref{fig:1D_jdiv}b. This current double layer generates a strong magnetic field at the rear target surface (Fig.~\ref{fig:1D_jdiv}c). 

The localized production of electrons near the target surface and the initial magnetization of the target are both crucial to the high amplitude surface magnetic field generation. Such a large surface field is produced because electrons undergo cyclotron rotation during the course of their transit of the target. This will be shown directly by the estimate for the rear surface magnetic field we construct in the following Section.


\section{1D scaling of rear surface magnetic field} \label{sec:1Destimates}

In this Section, we obtain a qualitative picture for cyclotron rotation-mediated surface field generation. We will additionally demonstrate the robustness of the field generation mechanism in 1D simulations to the choice of laser intensity, target thickness, and the strength of the seed magnetic field and predict an optimum range for field generation. Over a large range of parameters, we find that the surface field is well-predicted by a simple scaling.

The rear surface return current arises to screen the target from the hot electron current in the sheath and has approximately equal magnitude to the sheath current.
We estimate the sheath current density as
\begin{equation} \label{eqn:japprox}
    j_y \sim -|e| v_y n_{s},
\end{equation}
where $n_{s}$ is the number density and $v_y$ is the average $y$-directed velocity of hot electrons entering the sheath. $v_y$ is produced by the rotation of the electron momentum during the transit of the target. Given that the magnetic field within the target remains approximately equal to the applied field ($B_z \approx B_{\mathrm{seed}}$), and assuming the electrons are relativistic with longitudinal ($x$-directed) velocity $v_x \sim c$, the transverse velocity is approximately
\begin{equation} \label{eqn:vyapprox}
    v_y \sim v_x \sin{\left(\dfrac{\omega_{c0}}{\gamma} \Delta t\right)} \sim \dfrac{|e| B_{\mathrm{seed}} \Delta x}{\gamma m_e c},    
\end{equation}
where $\omega_{c0} \equiv |e|B_{\mathrm{seed}}/m_e c$ is the non-relativistic cyclotron frequency associated with $B_{\mathrm{seed}}$, $\Delta x$ is the target thickness, $\gamma$ is taken as a characteristic value for the hot electrons, and we have assumed the overall momentum rotation is small ($\omega_{c0}\Delta t/\gamma \ll 1$).

We now estimate the magnetic field generated by this current. The sheath size is approximately given by the electron Debye length, $\lambda_{De} \equiv \sqrt{T_e/4 \pi e^2 n_{s}}$. For a relativistic plasma, we have $T_e \approx (\gamma -1) m_e c^2$, which we substitute in the Debye length to give $\lambda_{De} \approx \sqrt{(\gamma-1) m_e c^2/4 \pi e^2 n_{s}}$. Approximating the current density as constant over $\lambda_{De}$, the magnetic field generated at the target surface is approximately $B_{\mathrm{gen}} \sim 4 \pi j_y \lambda_{De}/c$. Combining this with Eqs.~(\ref{eqn:japprox}) and~(\ref{eqn:vyapprox}), the magnetic field generated at the rear target surface can be approximated as
\begin{equation} \label{eqn:Blambda_D}
    \dfrac{B_{\mathrm{gen}}}{B_{\mathrm{seed}}} \sim - \sqrt{\dfrac{4 \pi \left(\gamma-1\right) e^2 n_{s} \Delta x^2}{\gamma^2 m_e c^2}} \approx - \dfrac{\Delta x}{\lambda_{De}},
\end{equation}
where the last expression assumes the plasma is sufficiently relativistic that $\gamma-1 \approx \gamma$.

The surface magnetic field generation is inherently a kinetic effect and can be thought of as an overshoot of the diamagnetic effect. This can be seen directly through an alternate approach to deriving this equation.
The diamagnetic effect occurs when charged particles undergo cyclotron motion in a magnetic field which results in a net current that acts to reduce the field. Normally, the cyclotron motion and the net current are co-located, i.e. the rotation of the charged particles occurs in the same spatial region as the net current. This is the only possibility if the plasma is described as a single fluid in lieu of a kinetic description.
However, the target we consider is conductive and inhibits changes to the embedded magnetic field. Although hot electrons undergo rotation in the target, they are only able to generate a magnetic field in the sheath.
This magnetic field grows until the plasma-generated field in the sheath is able to undo the momentum rotation of electrons transiting the target, based on which we expect 
\begin{equation} \label{eqn:omega_bal}
    \dfrac{\omega_{c1} \Delta t_s}{\gamma} \approx \dfrac{\omega_{c0} \Delta t}{\gamma},
\end{equation}
where $\omega_{c1}$ is the non-relativistic cyclotron frequency associated with $B_{\mathrm{gen}}$ and $\Delta t_s$ is the time the electron spends in the sheath. 
Assuming the electron motion is relativistic and $\Delta t_s \sim \lambda_{De}/c$, Eq.~(\ref{eqn:omega_bal}) gives the same result as Eq.~(\ref{eqn:Blambda_D}).
This analysis also confirms what we stated at the end of Sec.~\ref{sec:1Ddemo}: the localized production of hot electrons at the front target surface and the embedded magnetic field are both key to producing a strong rear surface field.

\subsection{Estimate for sheath density in a laser-irradiated target} \label{sec:n_s}

As written, Eq.~(\ref{eqn:Blambda_D}) involves the sheath density $n_s$ and the characteristic hot electron $\gamma$-factor, both of which should in principle be measured from simulations. However, to obtain a simple predictive scaling, we now specifically consider the case of a short scale length preplasma (scale length $<$ laser wavelength) and a reasonably short laser pulse ($\sim 100$~fs). 
Under these conditions, we estimate the sheath density as roughly $n_{s} \sim \gamma n_{cr} \sim a_0 n_{cr}$.

The origin of this estimate can be seen straightforwardly by considering the transfer of laser energy into hot electrons in the short scale length preplasma. The maximum number of electrons the laser can interact with and accelerate in half a laser cycle can be estimated from the condition where the laser transfers a substantial fraction of its energy to electrons. This energy balance is given by
\begin{align} \label{eqn:energy}
\begin{split}
    \left(\gamma -1\right)m_ec^2 N & \simeq \dfrac{c}{8\pi \omega_0} \int_0^\pi \left( E^2 + B^2 \right) \mathrm{d}\left(\omega_0 t\right) \\
    & =  \dfrac{E_0^2 c}{8 \omega_0} = a_0^2 \dfrac{m_e^2 c^3 \omega_0}{8 e^2},
\end{split}    
\end{align}    
where $N$ is the number of electrons the laser accelerates per unit area during the half-cycle and $a_0 \equiv |e| E_0/m_e c \omega_0$ is the normalized vector potential for the laser pulse with maximum amplitude $E_0$ and frequency $\omega_0$. 
The maximum number density of hot electrons streaming through the target into the sheath is thus approximately
\begin{equation} \label{eqn:energy_general}
 n_s \lesssim 2 N/\beta \lambda_0 = \dfrac{a_0^2}{\beta \left(\gamma-1\right)}\dfrac{n_{cr}}{2},
\end{equation}
where we have divided $N$ by $\beta \lambda_0/2$ with $\beta = v/c$ to approximate the hot electrons being distributed within the target over the full half-cycle. This may introduce an underestimate for the density as the electrons are often observed to be more strongly bunched (for example, in Ref.~\citenum{wilks1992pond_scale}).

For a sufficiently short laser pulse and preplasma scale length, it is well established that the electron energy roughly follows the ponderomotive scaling regardless of the exact acceleration mechanism~\cite{kruer1985jxB,wilks1992pond_scale,beg1997PoP_begscaling,lefebvre1997absorption}. In the ponderomotive limit, and the limit where $a_0^2 \gg 1$, we therefore have
\begin{align} \label{eqn:n_s}
    \begin{split}
    \gamma & \approx \sqrt{1+a_0^2} \approx a_0 \\
    n_s & \lesssim \dfrac{\sqrt{1+a_0^2}}{\sqrt{1+a_0^2}-1} \dfrac{a_0 n_{cr}}{2} \approx \dfrac{a_0 n_{cr}}{2}.
    \end{split}
\end{align}

This estimate is consistent with the expectation that the laser interacts with electrons in a preplasma up to the relativistically adjusted critical density surface, where $n = \gamma n_{cr} \approx a_0 n_{cr}$. Eq.~(\ref{eqn:energy_general}) is not specific to normal incidence or highly relativistic motion.

\subsection{Scaling and limit on maximum generated field strength}

For the case of a short scale length preplasma and a reasonably short laser pulse, we therefore take as an order-of-magnitude estimate $n_s \sim a_0 n_{cr}$ (equivalently, $n_s (\gamma - 1)/\gamma^2 \sim n_{cr}$) in Eq.~(\ref{eqn:Blambda_D}), based on which we expect the strength of the generated magnetic field to scale as roughly
\begin{equation} \label{eqn:Bscaling}
    \dfrac{B_{\mathrm{gen}}}{B_{\mathrm{seed}}} \sim - \sqrt{\dfrac{4 \pi e^2 n_{cr} \Delta x^2}{m_e c^2}} = - \dfrac{2\pi \Delta x}{\lambda_0}.
\end{equation}

\begin{figure}
    \centering
    \includegraphics[width=0.7\linewidth]{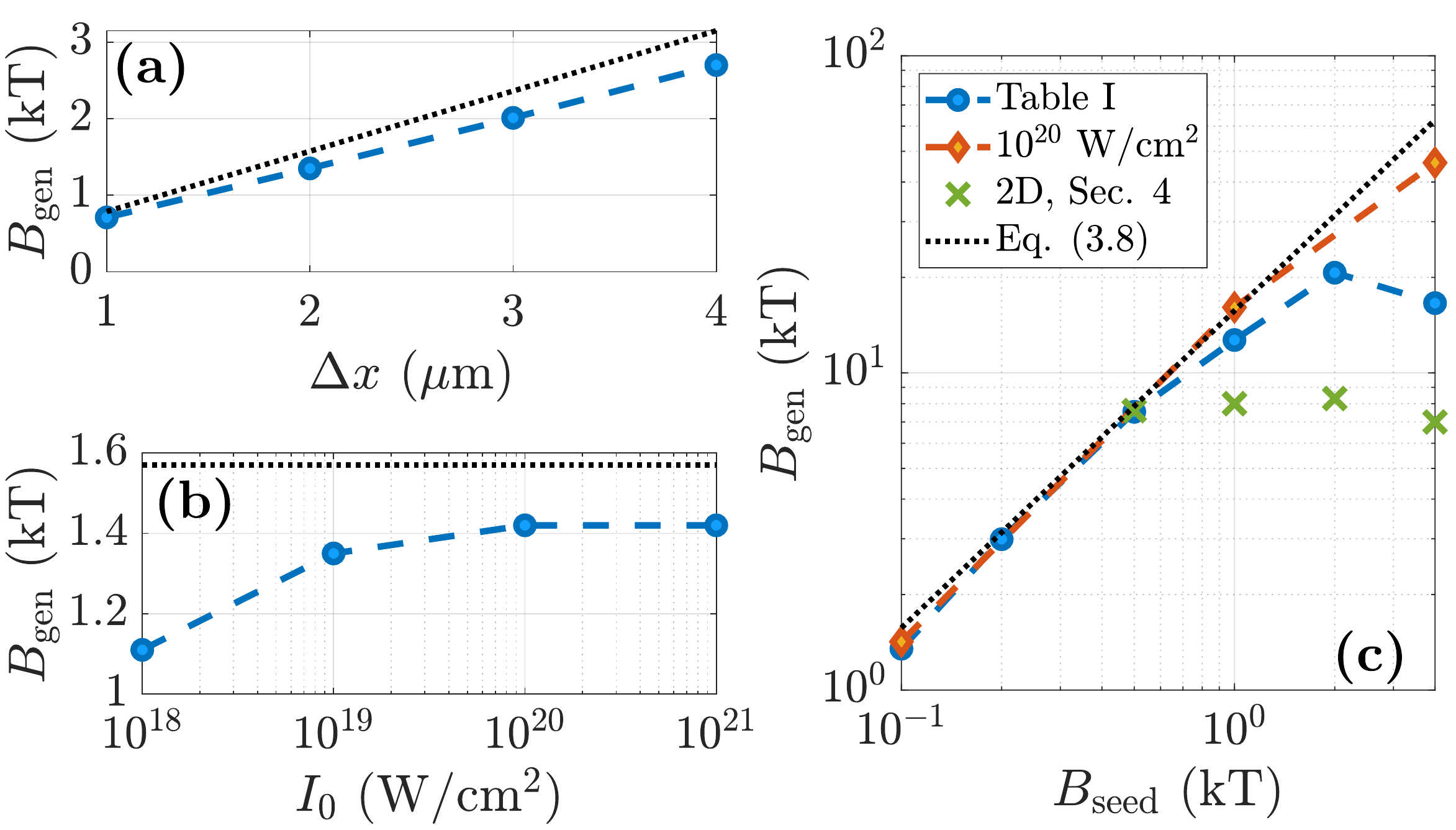}    
    \caption{Maximum rear surface magnetic field in 1D parameter scans. (a) Scan over target thickness. (b) Scan over peak intensity. (c) Scan over seed magnetic field strength.
    The simulation parameters not scanned over are as given in Table~\ref{table:1DPIC}. The black dotted lines correspond to Eq.~(\ref{eqn:Bscaling}). 
    }
    \label{fig:1Dscan}
\end{figure}

We now additionally estimate the maximum surface magnetic field which can be produced. While Eq.~(\ref{eqn:Bscaling}) provides a good prediction of the generated magnetic field strength over a wide range of conditions (see Fig.~\ref{fig:1Dscan}), this scaling breaks down if the seed magnetic field is sufficiently strong for electrons to undergo a significant fraction of a cyclotron rotation within the target. 
We roughly estimate the maximum surface magnetic field which can be produced by estimating $v_y \sim c$, which occurs when the target thickness is equal to the Larmor radius $\rho_e \equiv c p_x/|e|B_\mathrm{seed}$. Estimating $c p_x \sim \sqrt{\gamma^2-1} m_e c^2$ and setting $\rho_e = \Delta x$ gives
\begin{equation} \label{eqn:B0max}
    B_{\mathrm{seed}}^* \sim \dfrac{\sqrt{\gamma^2 -1} m_e c^2}{|e| \Delta x} \approx \dfrac{a_0 m_e c^2}{|e| \Delta x},
\end{equation}
where we have approximated $\sqrt{\gamma^2-1} \approx a_0$ as discussed in Sec.~\ref{sec:n_s}.
The maximum amplitude of the magnetic field that can be generated is roughly (employing $v_y\sim c$ in Eq.~(\ref{eqn:japprox}) and retaining $B_{\mathrm{gen}} \sim 4 \pi j_y \lambda_{De} /c$),
\begin{equation} \label{eqn:B1max}
    B_{\mathrm{gen}}^* \sim - \sqrt{4 \pi \gamma n_s m_e c^2} \approx - a_0 \sqrt{4 \pi n_{cr} m_e c^2}.
\end{equation}

For a 0.8~$\mu$m laser wavelength and a 2~$\mu$m thick target, we therefore predict the maximum magnetic field amplitude that can be generated to be $B_{\mathrm{gen}}^* \approx 13\, a_0$~kT occurring at an initial seed amplitude of $B_{\mathrm{seed}}^* \approx 0.85\, a_0$~kT.
Due to the nature of the estimate we performed, Eqs.~(\ref{eqn:B0max}) and~(\ref{eqn:B1max}) are undoubtedly overestimates, nevertheless, they establish the optimum seed magnetic field for surface field generation to be on the order of kT for few-$\mu$m-thick targets with $a_0 \lesssim 10$. Such fields are rapidly becoming experimentally relevant~\cite{santos2018coil,ivanov2018zebra}.

Below $B_{\mathrm{seed}}^*$, based on Eq.~(\ref{eqn:Bscaling}), we expect the plasma-generated magnetic field strength $B_{\mathrm{gen}}$ to be insensitive to the laser intensity, and to increase linearly with the target thickness and the seed magnetic field strength. 
Fig.~\ref{fig:1Dscan} shows how $B_{\mathrm{gen}}$ scales with these parameters.
Overall, we find good agreement between the predicted scaling and 1D PIC simulation results over a wide range of parameters, including in the approximate magnitude of $|B_{\mathrm{gen}}/B_{\mathrm{seed}}|$, which for the nominal case we predict to be $\sim 16$ based on Eq.~(\ref{eqn:Bscaling}) and observe in 1D PIC simulation to be 14-16.

The assumptions made to obtain Eq.~(\ref{eqn:Bscaling}) break down if the electron motion becomes sub-relativistic.
Correspondingly, we find that the magnitude of the rear surface magnetic field is insensitive to laser intensity for $I_0 \gtrsim 10^{19}$~W/cm$^2$, but begins to drop below this threshold as the electron motion in the laser becomes less relativistic, as shown in Fig.~\ref{fig:1Dscan}b.

We also find that the magnetic field generation is reduced relative to the prediction of Eq.~(\ref{eqn:Bscaling}) as the seed magnetic field strength approaches $B_{\mathrm{seed}}^*$, corresponding to the regime where the electrons complete a noticeable fraction of a cyclotron rotation within the target. 
For the parameters given in Table~\ref{table:1DPIC}, when $B_{\mathrm{seed}} \gtrsim 1$~kT, the generated magnetic field begins to deviate from the predicted value based on Eq.~(\ref{eqn:Bscaling}). 
As shown in Fig.~\ref{fig:1Dscan}c, the maximum magnitude of the surface magnetic field is approximately 19~kT corresponding to a seed field of 2~kT. Our observed $B_{\mathrm{gen}}^*$ and $B_{\mathrm{seed}}^*$ agree with the predictions of Eqs.~(\ref{eqn:B1max}) and~(\ref{eqn:B0max}) to within a factor of 1.5.
For $B_{\mathrm{seed}} \gtrsim B_{\mathrm{seed}}^*$, the generated magnetic field is reduced relative to $B_{\mathrm{gen}}^*$.

As we have discussed, the strongest surface generated magnetic field is produced for $B_{\mathrm{seed}} \sim B_{\mathrm{seed}}^*$. In this regime, both the surface generated and the seed magnetic fields can have a notable and application-relevant effect on the plasma dynamics. In the following section, we consider the effect of the seed and surface-generated magnetic fields on a laser-irradiated target in 2D. For $B_{\mathrm{seed}} \gtrsim B_{\mathrm{seed}}^*$, the magnetic field can substantially alter target expansion and the associated ion acceleration.


\section{Plasma expansion with a strong applied magnetic field} \label{sec:trapping}

In this Section, we discuss the regime in which the seed and plasma-generated surface magnetic fields are sufficiently strong to affect the overall expansion of the laser-irradiated target.
For $B_{\mathrm{seed}} \gtrsim B_{\mathrm{seed}}^*$, electrons become trapped near the target surfaces, restricting the rear surface expansion and associated ion acceleration. At the same time, the front surface expansion is enhanced, increasing the energy of backward-accelerated ions. For sufficient $B_{\mathrm{seed}}$, the energy and number of ions accelerated backward by the expanding front surface can exceed those accelerated from the rear surface, an unusual situation for thin laser-irradiated targets with a preplasma~\cite{ceccotti2007ion_fwdbwd}.

In Sections~\ref{sec:1Ddemo} and~\ref{sec:1Destimates}, we conducted 1D simulations to illustrate the magnetic field generation process. However, 1D geometry neglects higher-dimensional effects such as the finite laser spot size which in more realistic simulations (e.g. 2D) leads to the generation of azimuthal magnetic fields~\cite{sarri2012azimuthal,nakatsutsumi2018azimuthal,huang2019bgen_solid} which could potentially compete with the $-z$-directed non-azimuthal surface magnetic field generation.

First, we demonstrate using 2D simulations that the surface magnetic field generation can disrupt the development of the usual rear-surface azimuthal field and can produce a stronger, non-azimuthal magnetic field at the rear target surface. Sufficiently strong azimuthal magnetic fields have been shown to impair ion acceleration via target normal sheath acceleration~\cite{nakatsutsumi2018azimuthal}. 
Second, we demonstrate that the presence of the seed magnetic field and the generation of the non-azimuthal surface field can exacerbate this effect. While the rear-surface expansion can be dramatically reduced, the front-surface expansion is enhanced and can even produce higher accelerated ion number than ordinary ($B_{\mathrm{seed}} = 0$) rear-surface target normal sheath acceleration (TNSA).

\subsection{Surface field generation in 2D simulations}

\begin{figure}
    \centering
    \includegraphics[width=0.7\linewidth]{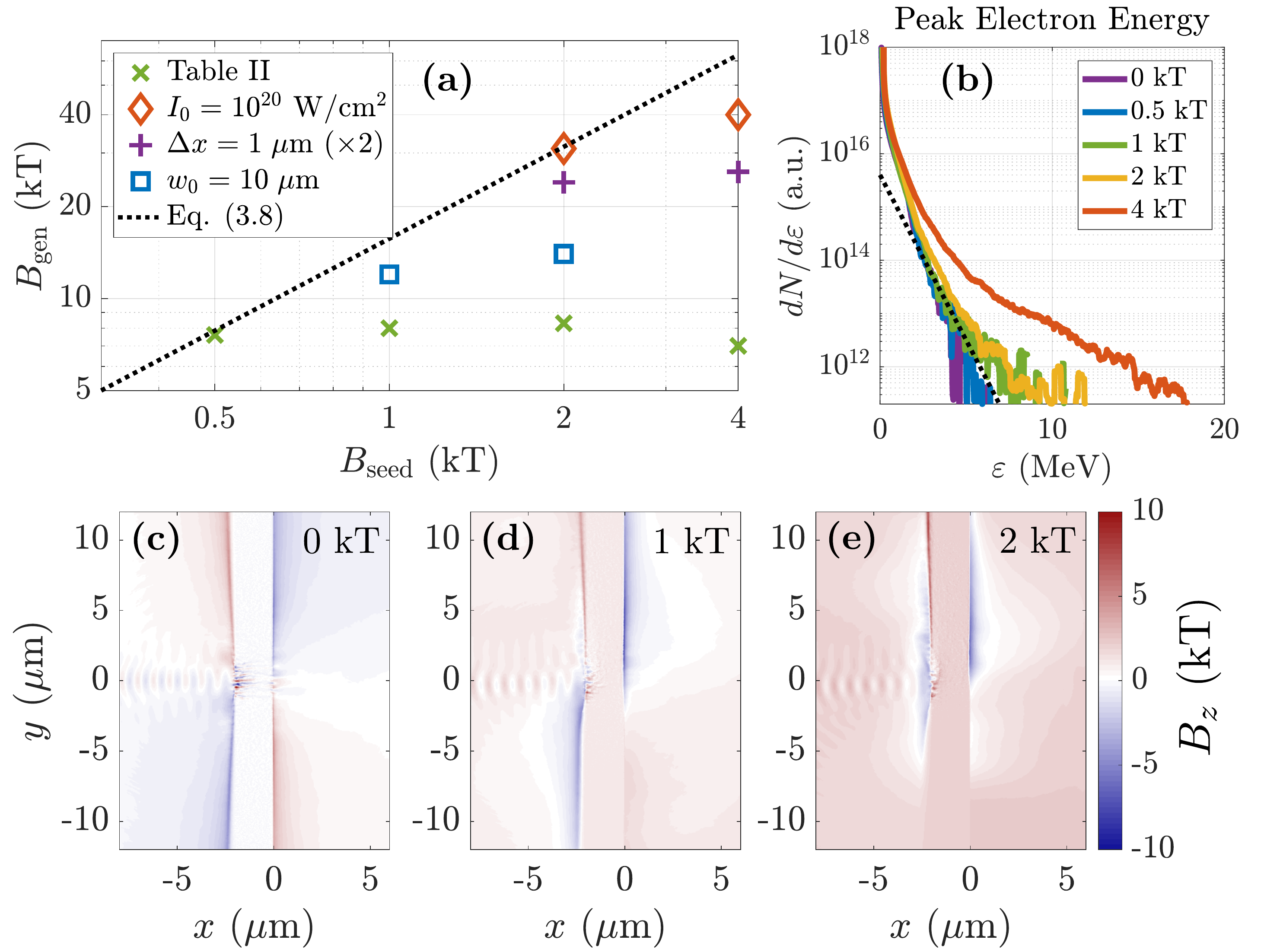}
    \caption{Surface magnetic field generation in 2D PIC simulations. (a)~Peak surface magnetic field in parameter scans. The legend indicates parameters which differ from the setup given in Table~\ref{table:2DPIC}. The results with $\Delta x = 1$~$\mu$m have been multiplied by a factor of 2 for the sake of comparison. (b)-(e)~Scan over $B_\mathrm{seed}$ for the conditions of Table~\ref{table:2DPIC}. (b)~Electron energy spectrum. Dotted line: pondermotive temperature $T_p=0.7$~MeV. (c)-(e)~Magnetic field profile at $t=75$~fs (c)~without an applied magnetic field, (d)~with $B_{\mathrm{seed}}=1$~kT, and (e)~with $B_{\mathrm{seed}} = 2$~kT.
    }
    \label{fig:2Dfield}
\end{figure}

We conduct 2D simulations with a finite laser spot size of 3~$\mu$m FWHM (Gaussian, electric field) and peak intensity $I_0 = 10^{19}$~W/cm$^2$. Additional parameters which differ from the 1D simulations of Secs.~\ref{sec:1Ddemo} and~\ref{sec:1Destimates} are given in Table~\ref{table:2DPIC}.
We begin with the case of $B_{\mathrm{seed}} = 0$ (no applied magnetic field). As the laser-heated electrons stream through the target, they generate an azimuthal field with maximum magnitude of approximately 3.8~kT (Fig.~\ref{fig:2Dfield}c). 
This field is associated with the outward radial streaming of electrons in the sheath~\cite{sarri2012azimuthal}.

\begin{table}
\centering
\begin{tabular}{ |l|l| }
  \hline
  \multicolumn{2}{|l|}{\textbf{Laser parameters} }\\
  \hline
  Spot size (Gaussian, electric field FWHM) & 3 $\mu$m\\
  \hline \hline
  \multicolumn{2}{|l|}{\textbf{Other parameters} }\\
  \hline
  Seed magnetic field ($\mathbf{B}=B_{\mathrm{seed}} \mathbf{\hat{z}}$) & $B_{\mathrm{seed}}=1$ kT \\  
  Spatial resolution & 50 cells/$\lambda_0$ \\
  Macroparticles per cell, electron and ion &  60 \\
  Size of simulation box ($x \times y$, $\mu$m) & $35 \times 70$ \\
  Time interval for averaging $B_z$ in figures & 20 fs \\
  \hline
  \end{tabular}
  \caption{Nominal 2D PIC simulation parameters with a planar target which differ from the 1D parameters given in Table~\ref{table:1DPIC}. The number of macroparticles per cell for the ions is increased to 120 within 0.2~$\mu$m of rear surface.}
  \label{table:2DPIC}
\end{table}

For $B_{\mathrm{seed}}>0$, the angular distribution of electrons entering the sheath is altered by the cyclotron rotation of electrons in the target. The magnetic field resulting from this offset becomes evident if $B_\mathrm{seed}$ is sufficiently large to produce $B_\mathrm{gen}$ at least comparable to the peak azimuthal magnetic field of the $B_\mathrm{seed}=0$ case. We can roughly estimate the minimum value of $B_\mathrm{seed}$ needed to produce a visible $B_\mathrm{gen}$ by considering the case when the cyclotron rotation in the target becomes comparable to the characteristic divergence angle $\alpha$ of hot electrons, i.e. $\rho_e \sin \alpha = \Delta x$. Typically, $\alpha \sim 25-45^\circ$ (e.g. Refs.~\citenum{sarri2012azimuthal,nakatsutsumi2018azimuthal}) such that $\sin \alpha \sim 1/2$, which implies that in general surface magnetic field generation will be observed in 2D and 3D geometry when
\begin{equation} \label{eqn:B0min_2D}
    B_\mathrm{seed} \gtrsim \dfrac{B_\mathrm{seed}^*}{2} \sim  \dfrac{a_0 m_e c^2}{2 |e| \Delta x},
\end{equation}
where $B_\mathrm{seed}^*$ is the seed field where we predict the generated field to be maximized (Eq.~\ref{eqn:B0max}).

The prediction of Eq.~\ref{eqn:B0min_2D} is in fairly good agreement with simulations. We observe that a seed field of at least 500~T is needed to substantially modify the surface field profile and increase the (negative) magnetic field amplitude relative to the $B_\mathrm{seed}=0$ case. As $B_\mathrm{seed}$ is increased, the azimuthal magnetic field at the rear target surface is suppressed and eventually overcome by the non-azimuthal surface field generation (e.g. Figs.~\ref{fig:2Dfield}d,e). 
As the field profile becomes more non-azimuthal, the generated magnetic field saturates at a peak value of approximately $-7$~kT for $B_\mathrm{seed} \gtrsim 500$~T (Fig.~\ref{fig:1Dscan}c). This value is roughly 2 times the peak magnetic field produced in the $B_\mathrm{seed}=0$ case. 
Although the saturation value is somewhat lower than what we observe in 1D simulations, $B_\mathrm{gen}$ is approximately the same at $B_{\mathrm{seed}} = 500$~T in 2D simulations as it is in 1D and is in good agreement with the prediction of Eq.~\ref{eqn:Bscaling} (see Fig.~\ref{fig:1Dscan}). 

The value of $B_\mathrm{seed}$ needed to modify the surface magnetic field and the saturation value of $B_\mathrm{gen}$ depends on the peak laser intensity, the target thickness, and to a lesser extent the laser spot size. We have conducted additional simulations with, separately, $I_0 = 10^{20}$~W/cm$^2$, $\Delta x = 1\;\mu$m, and $w_0 = 10\;\mu$m (Fig.~\ref{fig:2Dfield}a). The maximum amplitude of the azimuthal magnetic field produced in the $B_\mathrm{seed}=0$ case varies with these parameters due to changes in the hot electron population streaming through the rear target surface, as does the minimum $B_\mathrm{seed}$ required for the surface field to become non-azimuthal.
At the value of $B_\mathrm{seed}$ where $B_\mathrm{gen}$ becomes distinctly visible (left-most points in Fig.~\ref{fig:2Dfield}a), the generated surface field is in good agreement with the prediction of Eq.~\ref{eqn:Bscaling}.
In all cases, $B_\mathrm{gen}$ saturates at approximately this value, which is approximately 2 times the peak magnetic field produced with $B_\mathrm{seed}=0$. 
In the remainder of this work, we will analyze the case given in Table~\ref{table:2DPIC}.

In principle, the application of a seed magnetic field may also increase the electron energy if the magnetic field is sufficiently strong to rotate the electron momentum towards the laser polarization direction during direct laser acceleration~\cite{arefiev2020dla}, which could affect the generated surface field. 
However, for $B_\mathrm{seed}\lesssim 2$~kT, the seed field does not substantially change the number of accelerated electrons or the bulk of the electron energy spectrum (Fig.~\ref{fig:2Dfield}b).
For $B_{\mathrm{seed}} \lesssim 1$~kT, the hot electron temperature remains in good agreement with the ponderomotive scaling~\cite{wilks1992pond_scale} ($T_p = [(1+a_0^2)^{1/2}-1]m_e c^2 \approx 0.7$~MeV, black dotted line in Fig.~\ref{fig:2Dfield}b), and only a slight increase in the energy of the hottest part of the spectrum occurs for $B_{\mathrm{seed}} \leq 2$~kT.
As the seed magnitude is further increased to $B_{\mathrm{seed}}=4$~kT, the electron is substantially increased. However, at this seed amplitude the Larmor radius is smaller than the target thickness and electrons can be prevented from transiting all the way through the target (see Section~\ref{sec:expand}), and the surface field generation is actually reduced.

\subsection{Effect on target expansion and ion acceleration} \label{sec:expand}

\begin{figure*}
    \centering
    \includegraphics[width=0.95\linewidth]{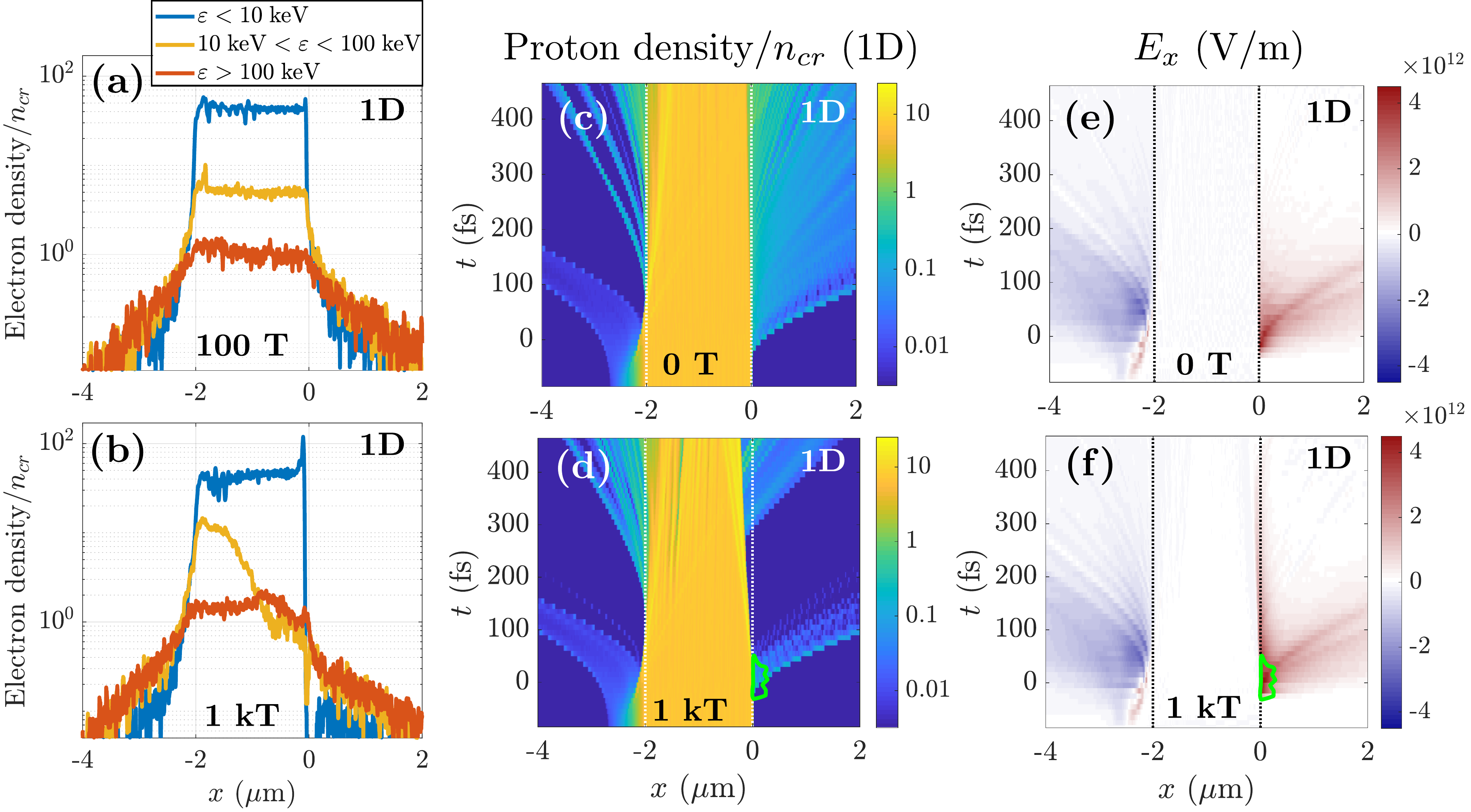}
    \caption{
    Modification of target expansion in 1D simulations by applied magnetic field. (a),(b)~Electron trapping near surfaces with (a)~$B_{\mathrm{seed}}=100$~T, and (b)~$B_{\mathrm{seed}}=1$~kT, at $t=175$~fs. 
    (c)-(f)~Target expansion with (c),(e)~$B_{\mathrm{seed}}=0$, and (d),(f)~$B_{\mathrm{seed}}=1$~kT. (c),(d) Proton density. (e),(f)~Electric field $E_x$. With $B_\mathrm{seed}=1$~kT, the rear surface expansion is terminated following the initial burst of ion acceleration once $|E_x|$ drops below $|B_z|$ (green contours in (d) and (f) denote where $|E_x|$ equals the maximum surface magnetic field magnitude at that time). Dotted lines in (c)-(f) denote initial target position.
    }
    \label{fig:expansion}
\end{figure*}

In the regime we are considering, both the seed and surface-generated magnetic fields can become sufficiently strong to inhibit the transport of electrons, resulting in electron trapping near the target surfaces and altering the target expansion and ion acceleration process.
To illustrate the effect of this trapping on the hot and return current electrons, we divide electrons into three populations based on their energy. 
For convenience, we perform this analysis based on 1D simulations.
When the applied magnetic field is below the kilotesla level (e.g. $B_{\mathrm{seed}}=100$~T in Fig.~\ref{fig:expansion}a), the electrons in all three energy bins become uniformly distributed throughout the target. However, with a 1~kT applied field (e.g. Fig.~\ref{fig:expansion}b), only the high energy electrons ($\varepsilon>100$~keV) become uniformly distributed. 
Electrons in the low energy bin ($\varepsilon<10$~keV) show a significant buildup at the rear target surface, while those in the middle energy bin (10~keV$<\varepsilon<100$~keV) are trapped near the front surface.

While the front surface trapping can be attributed to the strong seed magnetic field preventing the transit of moderate energy electrons through the target (which have a Larmor radius comparable to the target thickness), the rear surface buildup of electrons is more surprising. 
In the usual target normal sheath acceleration case with $B_{\mathrm{seed}}=0$, the transit of hot electrons through the target and compensating return current lead to the development of a thin ion-dominant layer at the rear target surface.
With no applied magnetic field or with a weak applied magnetic field, this ion layer is rapidly accelerated by the hot electron sheath, quickly reducing, but not entirely eliminating, the charge separation and associated electric field (Fig.~\ref{fig:expansion}e).
In this case, hot electrons remain free to transit the target and provide a continual acceleration of ions from the rear target surface as the target expands (e.g. Fig.~\ref{fig:expansion}c).

However, when the seed magnetic field is sufficiently strong ($B_{\mathrm{seed}}\gtrsim 1$~kT), the plasma-generated magnetic field in the rear sheath becomes comparable in strength to the sheath electric field (green contour in Fig.~\ref{fig:expansion}f), substantially altering the motion of electrons within the sheath and terminating the acceleration of ions from the target (Fig.~\ref{fig:expansion}d). This leads to a maintained ion density spike at the rear target surface, which eventually acts to attract the surrounding cold electrons, producing the density spike in the cold electron population seen in Fig.~\ref{fig:expansion}b.

This surface trapping has several consequences. 
First, from a modeling perspective, the localized production of electrons near the front target surface is a critical component of accurately modeling the surface trapping. Care must be taken in simulations of target expansion which substitute hot electrons for the laser-plasma interaction~\cite{welch2006ions_hote,kim2018llnl_ions} to account for this spatial localization.
We have demonstrated in Section~\ref{sec:1Destimates} that the initial spatial localization of hot electrons plays a substantial role in the generation of strong, asymmetric surface magnetic fields, and in this section we have shown that electrons do not eventually become uniformly distributed through the target in the presence of a strong seed field.

\begin{figure}
    \centering
    \includegraphics[width=0.9\linewidth]{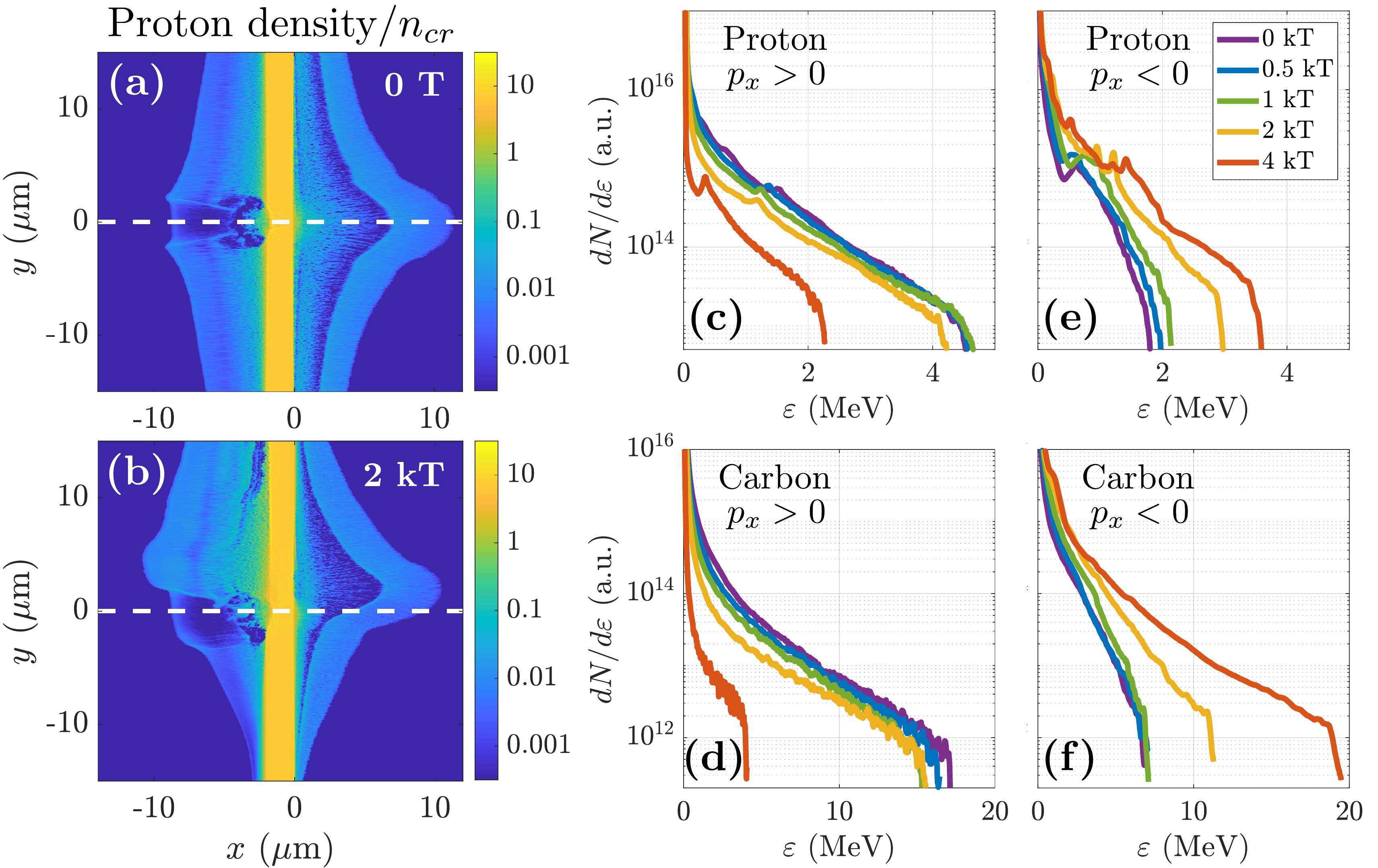}
    \caption{
    Modification of target expansion and ion energy in 2D PIC simulations by applied magnetic field. (a),(b)~Proton density at $t=430$~fs for (a)~$B_{\mathrm{seed}}=0$, and (b)~$B_{\mathrm{seed}}=2$~kT. The dashed line indicates the laser axis. Ion energy spectra for (c),(e) protons and (d),(f) carbon ions. (c),(d) Ions accelerated in the $+x$-direction, corresponding to ordinary rear-surface TNSA. (e),(f) Ions accelerated in the $-x$-direction from the laser-irradiated surface. Spectra were evaluated at $t=430$~fs; the cutoff energy changes by less than 10\% over the preceding 100~fs for all cases.
    }
    \label{fig:2Denergy}
\end{figure}

Second, the termination of target expansion as the sheath magnetic field begins to dominate over the electric field can substantially reduce the energy of ions accelerated from the rear surface.
As shown for our 2D simulations in Figure~\ref{fig:2Denergy}c,d, both the peak energy and the total number of accelerated ions with momentum $p_x > 0$ are strongly impacted by adding a seed magnetic field of $B_{\mathrm{seed}} \gtrsim 2$~kT.

At the same time, we observe a substantial increase in the energy and number of ions accelerated from the front surface ($p_x < 0$; Fig.~\ref{fig:2Denergy}e,f). This increased ion acceleration is attributable to the trapping and deflection of moderate energy electrons near the front surface. Ordinarily, i.e. with $B_{\mathrm{seed}} = 0$, ion acceleration from the front surface is suppressed by the radiation pressure exerted by the laser pulse, 
which initially causes front surface ions to be drawn into the target. This visibly digs a hole in the accelerated ion density on the laser axis (Fig.~\ref{fig:2Denergy}a). 
However, for $B_{\mathrm{seed}} > 0$, the hot electron cloud formed by the laser pulse is deflected transversely away from the laser axis and can accelerate ions from outside the laser spot (Fig.~\ref{fig:2Denergy}b).
This deflection, combined with the electron trapping near the front surface (e.g. Fig~\ref{fig:expansion}b) enhances the ion acceleration from the front surface.
For the 4~kT seed field, the ion energy may also be increased by the increased electron energy (see Fig.~\ref{fig:2Dfield}b). In this case, the carbon energy and number are enhanced beyond the $B_{\mathrm{seed}} = 0$ value (Fig.~\ref{fig:2Denergy}f).

\section{Summary} \label{sec:summary}

We have shown that laser-irradiated targets with an embedded target-transverse magnetic field do not behave purely diamagnetically when the laser is relativistically intense, but are instead able to generate strong surface magnetic fields. These surface magnetic fields result from the cyclotron rotation of the laser-heated and cold electron populations within the target and are fundamentally linked to the spatial localization of hot electron production by the laser pulse. This mechanism is robust over a range of laser and target parameters and produces surface field strengths on the order of 10-15 times the seed strength. 
We have formulated a simple predictive scaling in good agreement with both 1D and 2D particle-in-cell simulations, $B_\mathrm{gen} \sim - 2 \pi B_\mathrm{seed} \Delta x/\lambda_0$, and have demonstrated the relevance of surface field generation to applications. 
The applied seed and surface-generated surface fields can enact substantial electron trapping and visibly reduce and increase accelerated ion energies from the rear and front target surfaces, respectively. Both the changes in ion energy and the fields generated in these configurations may be experimentally visible, offering a potential route to experimental verification.

\section{Acknowledgements}

This research was supported by the DOE Office of Science under Grant No. DE-SC0018312.
Particle-in-cell simulations were performed using EPOCH~\cite{arber2015epoch},  developed under UK EPSRC Grant Nos. EP/G054940, EP/G055165, and EP/G056803.
This work used HPC resources of the Texas Advanced Computing Center (TACC) at the University of Texas at Austin and the Extreme Science and Engineering Discovery Environment (XSEDE)~\cite{towns2014xsede}, which is supported by National Science Foundation grant number ACI-1548562.
Data collaboration was supported by the SeedMe2 project~\cite{chourasia2017seedme} (http://dibbs.seedme.org).

\input{output.bbl}

\end{document}

%% file: output.bbl
%